# Spatial separation effect of asteroids with different albedos

Anatoly Kazantsev

Astronomical Observatory of Kyiv National Taras Shevchenko University, Ukraine

ankaz@observ.univ.kiev.ua

**ABSTRACT**

Numerical calculations of orbit evolutions of 1694 numbered asteroids included in the IRAS catalogue, from 13.11.1996 to 06.03.2006 were carried out. The values *da* – differences between the catalogue semimajor axes at 06.03.2006 and the calculated ones were computed. The average dependence *da* on albido *p* shows decrease of *da* at increase of *p*, and it is significant. In other words, semimajor axes of low-albedo asteroids are, on average, increasing as compared with high-albedo ones. Speed of such possible spatial separation for very dim and very bright asteroids of from 10 to 50km in order of magnitude is about 1 AU per $10^8$ years.

To explain this fact it may suppose an existence possibility of a non-gravitational effect. Such supposition is confirmed by distributions *p(a)* for asteroid families, above all, Flora family.

An analysis of errors and residuals in the used asteroid catalogues is evidence of such supposition.

## 1. Introduction

The influence of non-gravitational effects on evolution of asteroid orbits attracts more and more attention lately. It was seemed earlier, that the similar effects are negligible with respect for such large bodies as asteroids. However, already at present it is underlined in a number of publications the Yarkovsky effect and its analogues, acting during long period, can play an important role in orbit evolutions of kilometer-size asteroids and even larger (Nesvorný David and Bottke William F. 2004). At that, such effects cause: (a) drifting bodies to resonances, that causes replenishment of MEAs numbering (Bottke William F. et. al. 2006); (b) change spin rates and axis orientation of asteroids (Harris Alan W. and Pravec Peter 2006); (c) change ranges of semimajor axes of family asteroid orbits (Bottke William F. et. al. 2001). It is underlined in some papers that sometimes non-gravitational asteroid drifting occurs faster, than it predicted by theoretical estimations (Nesvorný David and Bottke William F. 2004). All these mean that the elaboration of influence of non-gravitational effects on evolution of asteroids and their orbits will be an urgent task still long time.

## 2. Some peculiarities of asteroid-albedo distribution

The conclusions of this article are largely based on the values of asteroid albedos. To date the most representative array of albedo values is the IRAS catalogue. More early publication of IRAS data (Tedesco et. el. 1992) contains albedos of 1884 asteroids. As the authors pointed, the data accuracy (sizes and albedos) averages 5 - 10%. The additional and revised data were published in 2002 (Tedesco E. F. et. al. 2002), where sizes and albedos of 2228 asteroids are presented.

To ground of some conclusions of the article it is necessary to consider one feature of asteroid albedos. It is a question of the well known dependence of asteroid albedos in the Main Belt *p* on semimajor axes of orbit *a*. The bodies with higher albedos are mainly situated in the internal zone of the MBA, and with lower albedos – in the external one.

The average dependence *p(a)* for MBAs, obtained from the IRAS catalogue, represented in Fig. 1. The dependences are represented for different inclinations and are approximated by a linear function. Here and later on to construction of linear dependences it will be used linear regress equation as

$$p = b_1 a + b_0 \quad (1)$$

Line 1 corresponds to range of inclinations from $0°$ to $5°$ (458 orbits), line 2 – the inclinations are large than $15°$ (416 orbits). One can see, the lines inclined to the axis *a* to a variable extent. I.e.

asteroids with less orbit inclinations have less differentiation of albedo values on average, than asteroids with larger orbit inclinations.

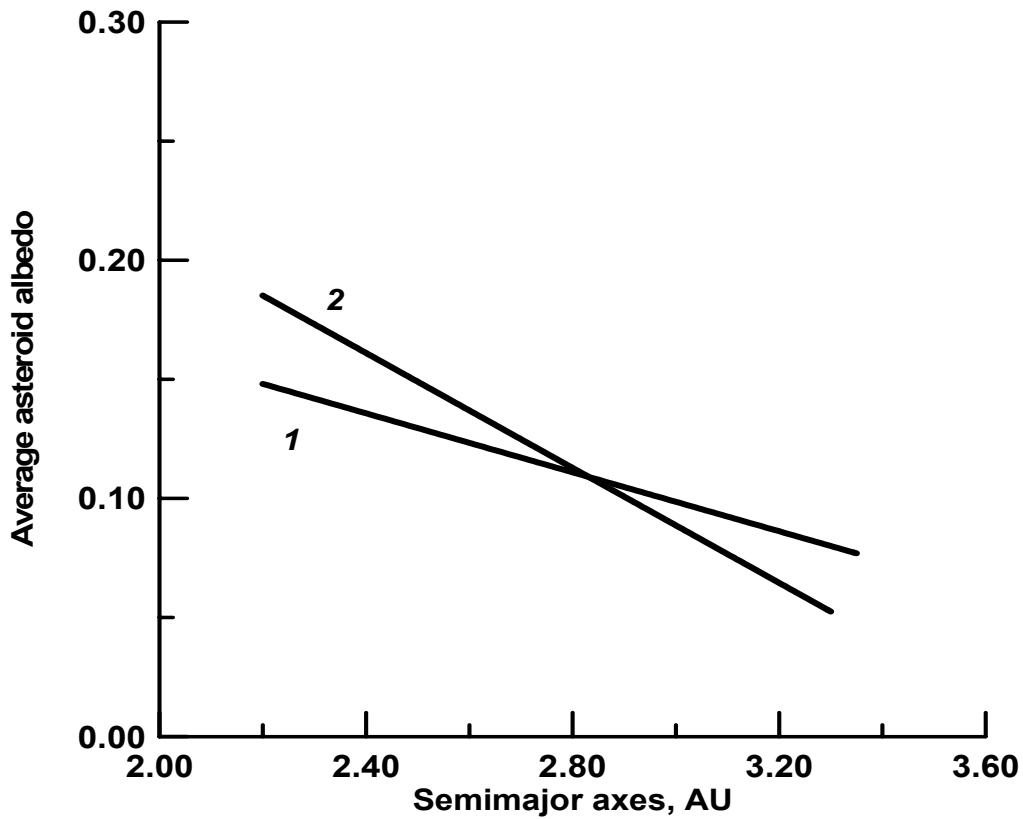

**Fig1** The average dependence *p(a)* for all asteroids of the Main Belt. 1- orbit inclinations i ≤ 5°, 2 – *i* > 15°

In principle, it can propose rather a simple and a logical explanation for this fact. The surfaces of asteroids in time are covered with layer of dust, generated during collisions of others bodies. At that, albedo of very bright surfaces must diminish, and pitch-dark ones – increase. A maximal concentration of dust must be nearby the ecliptic plane. Consequently asteroids on orbits with low inclinations must be covered with dust more quickly. That should be an explanation for dependences in Fig. 1.

Distributions for asteroids of from 10 to 50км will be analysed below. Therefore it is necessary to view dependences *p*(*a*) for asteroids with such sizes as well (Fig. 2). There are presented dependences for three ranges of orbit inclinations in the figure: 1 – from 0° to 5°, 2 – from 5° to 15°, 3 – greater than 15°. One can see from the Fig. 2 the increase of differentiation of asteroid albedos when increase of orbit inclinations has not casual, but systematic character. Coefficient $b_1$ for dependence 1 is almost three times less the coefficient for dependence 3. Such distinction of the dependences is significant in according to Kolmogorov-Smirnov test at level less 0.02. Aforecited explanation for change of dependence *p*(*a*) at changing of orbit inclinations (owing to covering asteroid with dust) is possible but not necessarily one and only. The main is the change of the dependence is really.

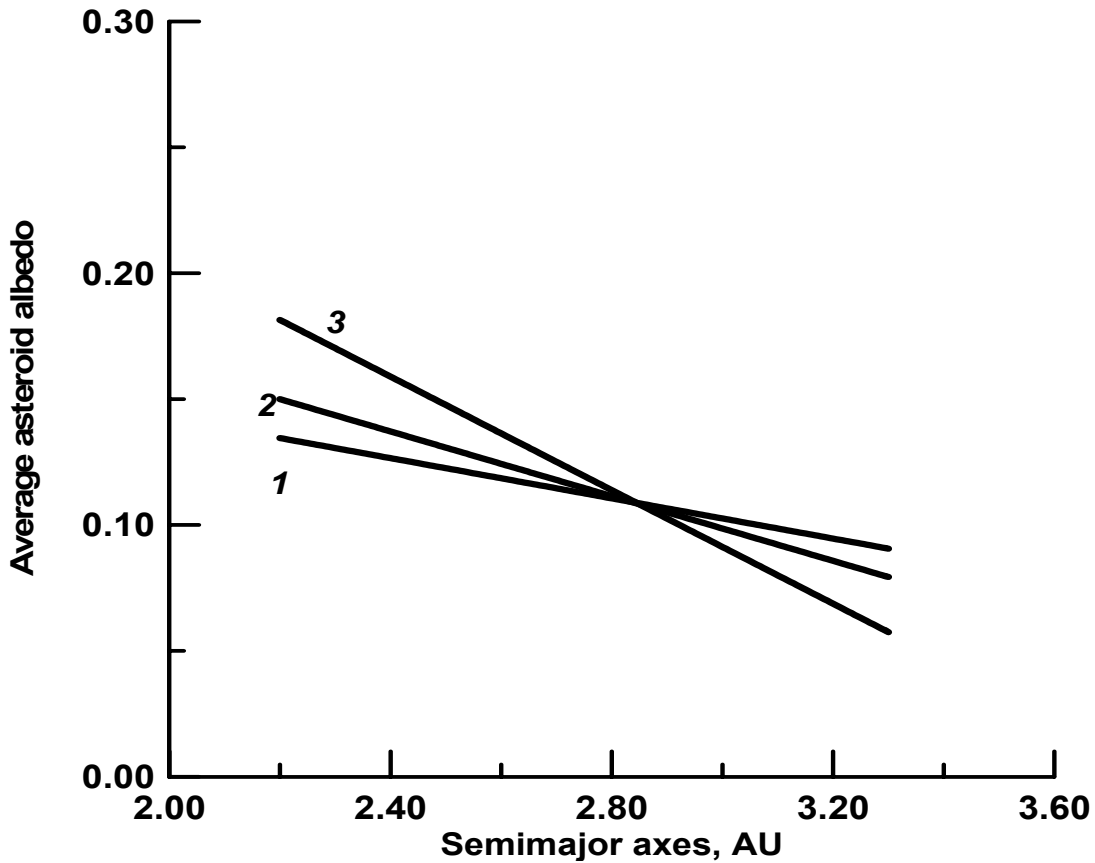

**Fig2** The average dependence *p(a)* for asteroids of from 10 to 50km. 1- orbit inclinations $i \leq 5°$, 2 – $i = 5°-15°$, 3 – $i > 15°$

### 3. Numerical calculations on revealing non-gravitational effects

The influence of non-gravitational effects can be estimated both analytically, and numerically. In the latter case the next approach is used. Asteroids with the well-determined orbits at some moment $t_0$ are selected. Then the numerical calculations of evolution of these orbits up to other moment in the future $t_1$ are carried out. The calculated orbit elements are compared with the catalogue ones at the same moment. The differences of elements can be caused by the several reasons, including an influence of non-gravitational effects as well.

Per se, such approach is statistical. It is clear that influence of non-gravitational effects on evolution of asteroid orbits is rather little. Besides, others factors have effect on evolution results of asteroid orbits, which practically cannot be estimated for a separate body. These are errors in catalogue orbit elements and possible influence of bodies unaccounted in the calculations (approaches with not very large asteroids or meteorite impacts). Under the analysis of orbit evolution of many bodies the similar influence can be essentially reduced and hence, more precisely can separate influences of non-gravitational effects.

Using such approach, in this paper it was made an attempt to reveal cumulative influence of non-gravitational effects on asteroid motions, first of all, on bodies of the Main belt. The calculations were carried out by numerical integration of equations of motion in rectangular coordinates by on the method described in the paper (Kazantsev A.M. 2002). There were taken into account the influence of 8 planets, Pluto, Ceres and two the largest asteroids (2 and 4).

The initial orbit elements of asteroids were taken from the year-book "Ephemeris of minor planets for 1996" (St. Petersburg. 1995). The initial moment was November 13, 1996. The integration of orbits was carried out to epoch of March 6, 2006 and the results were compared with the orbit elements, obtained in Institute of Applied Astronomy (IAA, St. Petersburg) by that date. These data were kindly rendered to us by prof. V.A.Shor. Besides the integration results were compared with the orbit elements, which are represented in the MPC catalogue at the same epoch (ftp://cfa-ftp.harvard.edu/pub/MPCORB/MPCORB.DAT).

As our task is revealing non-gravitational effects, it is important to know the sizes and albedos of the considered bodies. Therefore the asteroid orbits included to the IRAS catalogue (Tedesco Edward F. et. al. 2002), which contains the sizes and albedos of 2228 asteroids, were selected for the calculations. As a whole, number of MBA orbits included simultaneously both in the orbit catalogues and in the IRAS one consisted of 1694.

On completion of the numerical calculations, values *da* - differences between the semimajor axes of orbits in the IAA catalogue at March 6, 2006 and the semimajor axes of orbits obtained as a result of numerical evolution were determined. If *da* < 0, the calculated value of the semimajor axis is greater than the catalogue one.

It may note that almost all orbit elements in the IAA catalogue are completely coincide with correspondent ones in the MPC catalogue. It is clear, because in St. Petersburg the full observed database are used to calculate the orbit elements, as in MPC. Therefore, all given below dependences with values *da* are equally correct for both catalogues. Little differences between the catalogues are caused by different moments of the orbit calculations. The orbit elements in MPC catalogue have been calculated before the epoch of March 6, 2006, and the orbit elements in the IAA catalogue – after the epoch (in 2007). Therefore more observed data were used in the last catalogue.

The dependence of values *da* versus the asteroid sizes is represented in Fig. 3. For convenience the values *da* are designated X-direction. Discontinuity of *da* is caused by accuracy of the catalogue values of semimajor axes ($10^{-7}$ AU). Not of all orbits are represented in the figure, but only with $da < 2\times10^{-6}$ AU. It is clear, very large differences can't be caused by non-gravitational effects.

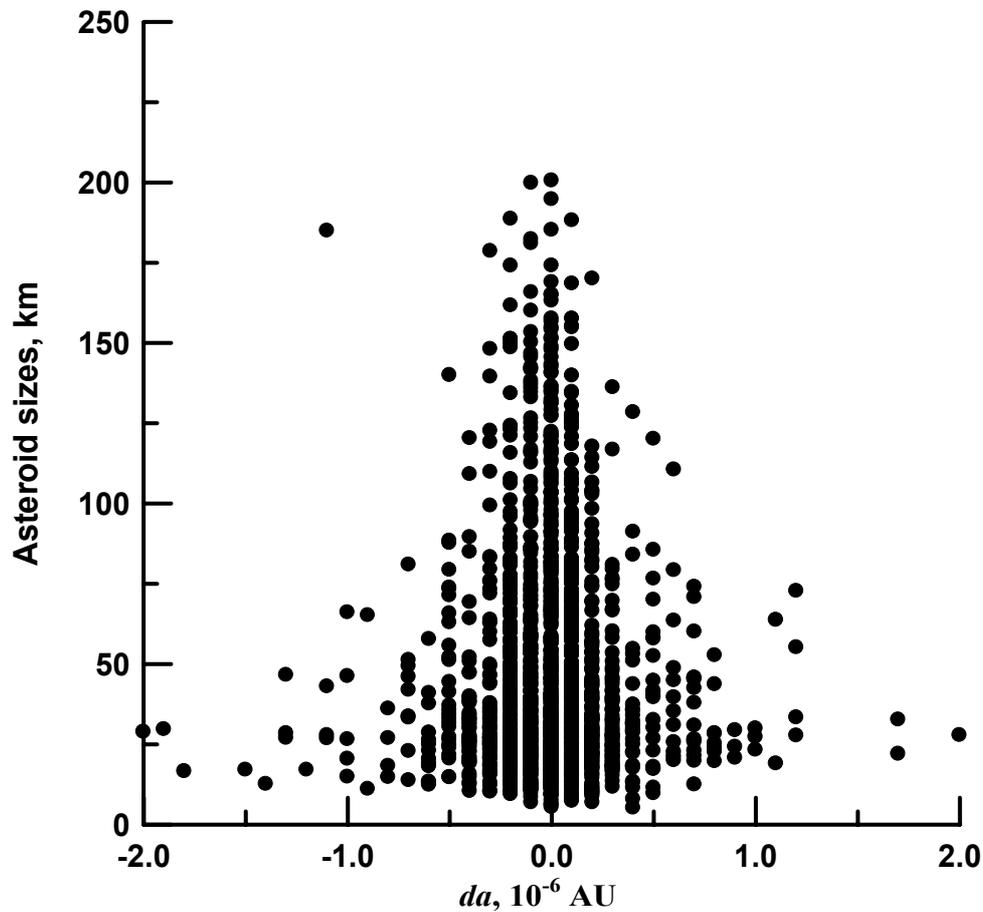

**Fig3** The dependence of differences between catalogue and calculated semimajor axes (*da*) versus the asteroid sizes

One can see from Fig. 3 the maximum number of orbits corresponds to zero differences (*da* = 0). So, at *da* = 0 number of orbits equal 392, at $da = -1\times10^{-7}$ – 303, at da = $1\times10^{-7}$ – 296. Under the difference increase the orbit number sharply decreases. And so, it is possible to make a conclusion that the accuracy of the method and the program of numerical calculations don't become apparent on differences *da*. If the program calculates precisely one orbit, all other orbits should be calculated precisely as well. As zero difference *da* corresponds to relative majority of orbits, all values *da*, which differ from zero should be caused by others factors. These may be: (a) inaccuracy of catalogue orbit elements, (b) influence of the bodies weren't taken into accounts, (c) cumulative influence of non-gravitational effects.

Our prime interest is to estimate the last factor. Its influence can to a greater degree become apparent on bodies having smaller sizes. However orbit elements of smaller asteroids are less exact. Such bodies have been opened later and they have faint brightness. Hence, during specification of theirs orbit elements from observations, rather large errors appear. Therefore there isn't a sense to look for manifestation of non-gravitational effects in the distribution *da(D)*.

It is clear, that for revealing a possible influence of the similar effects on the basis of others physical asteroid characteristics, it is necessary to exclude not only the smallest bodies, but the largest ones as well. Therefore for the further analysis the bodies with the sizes from 10km up to 50km were selected.

Besides, it is necessary to take into account that influence of non-gravitational effects on semimajor axes should be rather small, otherwise this influence would be already revealed. Consequently, it is hardly to explain appreciable absolute values *da* by non-gravitational effects. On the one hand, in order to reveal a possible non-gravitational effect it is necessary to select orbits with small values *da*. On the other hand, numbering of the selected orbits should be rather great so as to carry out statistic investigations. Therefore we have limited our sample in a range of values $da = \pm 4 \times 10^{-7}$ AU. There were 1018 such orbits.

As there isn't a possibility to reveal manifestation of non-gravitational effects in the distribution *da(D)*, it is possible to consider the dependence of *da* upon the asteroid albedos *p*. Such dependence for the full sample is shown in Fig. 4,*a*. Separate asteroid orbits are denoted by the points, dotted line – the average dependence. (In it and late on the average dependences are lined by least-squares method). The decrease of average value *da* under albedo increase may represent a certain interest if this dependence is significant. The average dependence *da(p)* is represented by linear regress equation

$$da = b_1 p + b_0 \qquad (2)$$

For the average dependence in Fig. 4,*a* the modulus of factor $b_1$ is only a little greater than its standard error. The values of the factor and its standard error are equal correspondingly –0.088 and 0.067. Any value to be significant from the point of view of the mathematical statistics, it should be at least twice as exceed its standard error. Therefore there is no a reason to speak about detection of any real dependence in this case.

In this connection, it is necessary to recollect the obtained above conclusion that asteroids with low orbit inclinations have less differentiation of albedo values on average, than asteroids with high orbit inclinations. Consequently, influence of some non-gravitational effect must stronger show up for asteroids with high orbit inclinations. Therefore, asteroid orbits with inclination $i > 10^o$ were selected for subsequent analysis.

The amount of residuary orbits was 436. The dependence *da(p)* for the new sample is represented in Fig. 4,*b*. The dotted line, as before, marks the average linear dependence. Though the orbit number was reduced more than twice, the standard error of the coefficient of regression became almost three times less than modulus of the coefficient. The values of the coefficient and its standard error are equal accordingly to -0.303 and 0.101.

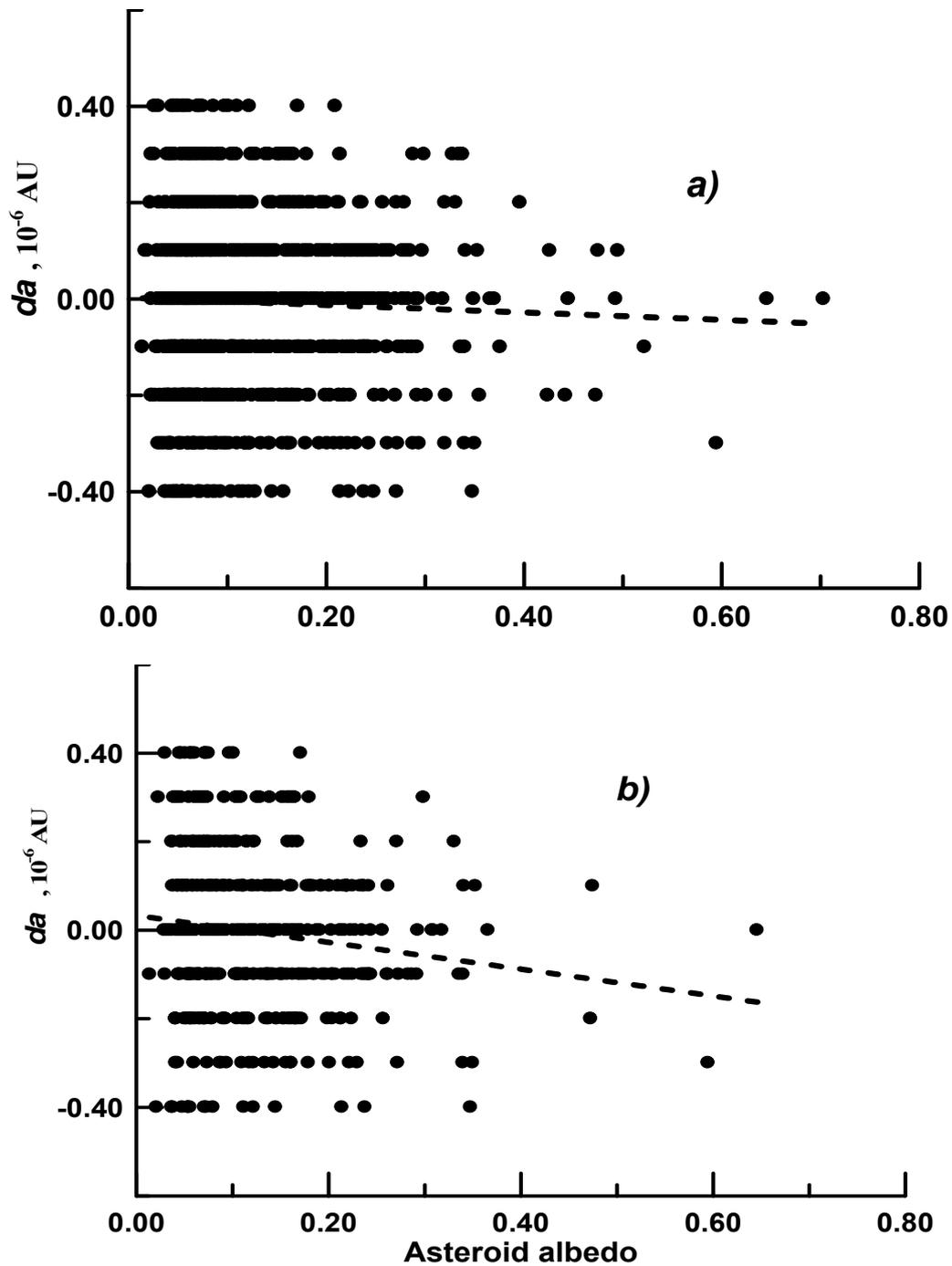

Fig4 Dependences of values *da* on asteroid albedos: a) for the full sample; b) for asteroids with orbit inclinations $i > 10°$

It may seem from Fig. 4,*b*, a significant role in these values play four orbits with albedo greater 0.4. The dependence *da(p)* without these four orbits gives the following values of coefficient of regression and its standard error: –0.326 and 0.11. If the minimal values of orbit inclination of the sample to increase to $i > 13^o$, the coefficient of regression becomes still more significant. The coefficient exceeds its error by a factor of 3.25, though the orbits numbering decreases up to 247.

The Fisher's test (F test) is generally used for determination of statistical significance of a linear regression. We have also applied this criterion to determination of significance of the dependence *p(a)*. It turned out, that both the dependence in Fig. 4,*b* and the similar dependence without four asteroids with the most high albedos are significant at a level less 0.005. Hence, could say the results of the performed numerical calculations indicate on existence of significant dependence of values *da* upon asteroid albedos. One of reasons of such result may be an action of some non-gravitational effect.

**4. A short consideration of orbit errors and residuals**

In principle, the dependence *da(p)* can be caused both by the real physical factors (action of some non-gravitational effects), and by quite other reasons. It is possible we deal with a specific influence of errors in asteroid catalogues. To a more concrete answer this question it is expedient to carry out a short consideration of orbit errors and residuals in the catalogues.

At once, it is necessary to mark that orbit errors and residuals – these are a few different values. Errors are differences between exact orbit elements and catalogue ones. Exact elements – these are very approximate to the catalogue ones, but unknown values. Exact elements exactly determine osculating orbit of concrete asteroid at a pointed epoch. Residuals – these are root-mean-square values of differences between observed positions of asteroid and calculated ones. At large residuals the correspondent observed positions are eliminated from the array which is applied to orbit determination. Therefore, very large residuals are absent in catalogues. In the catalogue at 1996 residuals lie within of $0.5'' - 6''.0$, in the catalogue at 2006 – from $0.5''$ to $1''.5$.

It is possible, at large residuals the large errors can take place, caused by precision of observed data. However, a principal possibility exists that at larger residuals less errors can occur.

In our case, for revealing a possible non-gravitational effect it is necessary as much as possible to exclude orbits with sizeable errors. On our opinion, values *da* may be analogues of orbit errors. As stated above, large values *da* are caused either by catalogue errors, or influence of the bodies weren't taken into accounts for the calculations. At any case, relatively small values *da* should correspond to more precise orbits. The dependence *da(p)* for $|da| \leq 0.4$ (Fig.4,*b*) is significant. For the similar sample of orbits (asteroid sizes from 10 to 50km, orbit inclinations higher $10^o$) but with all values *da*, the correspondent dependence *da(p)* is less distinct and insignificant. Ratio of coefficient of regression to its standard error is 1.02. Consequently dependence *da(p)* is more significant for more precise orbits. This fact may be an argument for supposition of existence possibility of non-gravitational effect also.

As to residuals, their connection with possible non-gravitational effect is quite different. One may remind that residuals – are differences between observed asteroid positions and the calculated ones. No non-gravitational effects are taken into account for obtained the calculated positions. Consequently, the real influence of some non-gravitational effect on motion of asteroids must results in increase in residuals. It means in our case, that if such effect exists, dependence *da(p)* should be more distinct for orbits with larger residuals. It is necessary to check up a possible influence of non-gravitational effects by the example of study asteroid orbits.

For that we will consider again the orbit sample, for which dependence *da(p)* is presented in Fig. 4,*b*. The sample includes 436 orbits. The correspondent rangers of residual are: at 1996 ($rms_{96}$) – from 0.50" to 3".30, at 2006. ($rms_{06}$) – from 0.51" to 1".28. More than fourth of values $rms_{06}$ does not exceed 0.56". From the sample four sub-samples with different ranges of residuals were formed. The first sub-sample includes orbits with minimum residuals: $rms_{96} \leq 1".5$, $rms_{06} \leq 0.56"$. Their amount was 78. The second sub-sample contains orbits without maximal residuals: $rms_{96} \leq 1".7$, $rms_{06} \leq 0.8"$. Orbit number of the sub-sample was 304. In the third sub-sample were put orbits without minimum residuals: $rms_{96} \geq 1".0$, $rms_{06} \geq 0.56"$. Number of orbit of the sub-sample was 247. The fourth sub-sample includes orbits with maximal residuals: $rms_{96} \geq 1".65$, $rms_{06} \geq 0.60"$. The orbit number – 81.

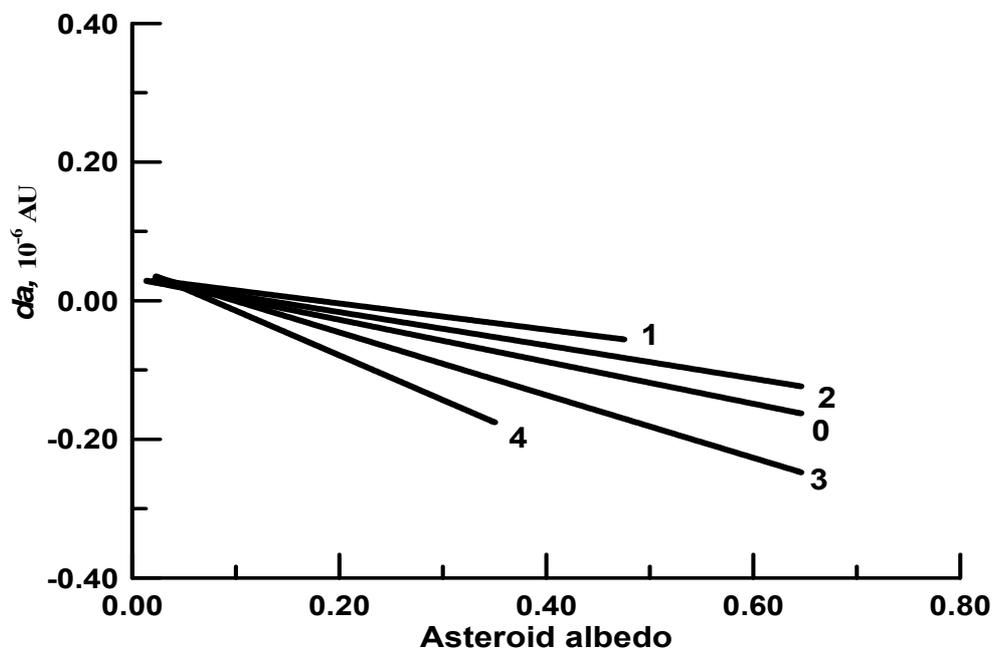

Fig5 The average dependences *da(p)* for sub-samples of different residuals.

Thus, we obtained four sub-samples with different ranges of residuals. The middle value of residuals increases at increase of the sub-sample number (1 – 4). It may add here the full orbit sample (436 orbits). The sample was denoted as zero sub-sample. The middle residual values of the full sample lie between correspondent values of 2 and 3 sub-samples.

Distributions *da(p)* were constructed for every sub-samples. Their average dependences are presented in Fig. 5. The dependence number corresponds to the sub-sample number. The orbit distributions aren't presented because they cover one another. For all dependences $da = 0$ at $p \approx 0.10$. One can see from the figure that dependences *da(p)* are more distinct for orbits with larger residuals. Absolutely values $b_1$ increase at increase of residuals. For dependences 0 and 3 levels of significance are less 0.005, for dependence 4 – is less 0.04. Data of sub-samples, residuals and coefficients $b_1$ are presented in Tab. 1.

Table 1 Residual rangers, asteroid numbering and values of coefficient of regress for all sub-samples

| $N_s$ | $rms_{96}$ | $rms_{06}$ | $N_{orb}$ | $b_1$ |
|---|---|---|---|---|
| 1 | $0''.50 - 1''.50$ | $0''.51 - 0''.56$ | 78 | −0.099 |
| 2 | $0''.50 - 1''.70$ | $0''.51 - 0''.80$ | 304 | −0.240 |
| 0 | $0''.50 - 3''.30$ | $0''.51 - 1''.28$ | 436 | −0.303 |
| 3 | $1''.00 - 3''.30$ | $0''.56 - 1''.28$ | 237 | −0.452 |
| 4 | $1''.65 - 3''.30$ | $0''.60 - 1''.28$ | 81 | −0.644 |

Increase slope of dependences *da(p)* at increase of residuals may regard as an argument in support of assumption of existence possibility of non-gravitational effect. Besides, this result points out that possible non-gravitational effect can partly influence on residual values in asteroid catalogues. Indeed, if non-gravitational effect has stronger action upon some asteroids, their orbits should be determine with larger residuals. And for such orbits dependences *da(p)* should have a steeper slope.

The represented results may cause a remark, if such non-gravitational effect exists, its influence during a long observation period could be estimated and improved set of orbital elements and residuals could be obtained. In this connection one might that the residuals are noticeably greater values in comparison with systematic orbit differences caused by possible non-gravitational effect. For an asteroid in MBA, residual 1 arcsec corresponds to uncertainty of 1000km in its space position. As to our estimations, the concerned non-gravitational effect causes, on the average, systematic changes in semimajor axis of asteroid orbit by $10^{-7}$ AU during 100 years. It changes the asteroid space position by noticeably less value than 1000km. Therefore, influence of such non-gravitational effect can't be revealed from observed data for a separate asteroid, and an improved set of orbital elements and residuals can't be obtained, especially if nobody supposes an existence possibility of the effect. The influence of the effect can be revealed by statistic way only, as it was used in the paper.

At the same time, chaotic changes of asteroid space position, caused by non-gravitational effect, may be grater than systematic ones and may affect residuals. Because dependence *da(p)* are more distinct for orbits with larger residuals.

It is necessary to mention errors in the IRAS catalogue. According to some publications (for example Lupishko D.F. 1996), the IRAS albedos contains noticeable errors. Comparison of IRAS albedos with occultation and with polarimetric ones have shown some systematic differences between the albedos. The IRAS albedos were corrected taking into account the differences. All calculations were repeated with the corrected albedos, and all results remained significant at the same level.

### 5. Revealing observational confirmations of possible non-gravitational effect

Criterion of the question decision can be the observed data. If to assume action of non-gravitational effects, they should become apparent on concrete distributions of the real asteroid orbits. As one can see from Fig. 4, semimajor axes of orbits of low-albedo asteroids should on average increase as compared with the semimajor axes of orbits of the high-albedo bodies. That is, a character distribution *p(a)* for asteroids with sizes of few tens kilometres should be take place.

In our opinion a confirmation of such effect it is necessary to search in the distributions *p(a)* of separate asteroid families (the families which exist not less than several millions years). On the basis of the performed calculations it is possible to estimate speed of spatial division for asteroids with different albedos. Probably, it is difficultly to say about any exact numerical estimation of the speed of spatial division for asteroids. It can estimate only the order of such speed value. As one can see from Fig. 4,*b*, the average difference of *da* for very dim asteroids (*p* = 0.03) and for very bright ones (*p*=0.40) is about $1 \times 10^{-7}$ AU. Consequently, the order of speed value of spatial division for dim and bright asteroids is 1 AU per 100 millions years. Zone of semimajor axes almost for any asteroid family is within a few hundredths of AU. It is universally recognized that the origination of asteroid families occurs as a result of break-up of a one large body. At that, asteroids of a single family obtain various albedos. It is clear, that the distribution of albedos upon semimajor axes for the recently formed family should be on average isotropic. If the sought non-gravitational mechanism is really acting, during several millions years the family asteroids should acquire an anisotropic distribution *p(a)*. Just that distribution, that average asteroid albedo decreases under increase of semimajor axes. Therefore we have made an attempt to test a reality of the sought non-gravitational mechanism by distributions $p(a)$ for separate asteroid families.

As is well known, the asteroid families are determined by on correspondent ranges of proper semimajor axes $a'$, proper eccentricities $e'$ and proper inclinations $i'$. As a matter of fact, proper elements are Keplerian elements corrected with regard to secular perturbations.

Apparently, the most complete catalogue of proper asteroid elements is the catalogue Milani and Knezevic located to the Internet - address http: // hamilton.dm.unipi.it/cgi-bin/astdys/. Just this catalogue was used in this paper to selection of family asteroids. To such selection it is necessary to know the rangers of proper elements for the families. Different authors give some different ranges of proper elements for the same families. However, there aren't essential distinctions between these data. Such distinctions are mainly caused by asteroid sample, used for determination of ranges. The more numerous sample the more exact ranges. To our task we used publication Zappala, V. and Cellino A. (1994). In this paper on sample about 4000 orbits the ranges of proper elements for 26 asteroid families had been determined. As the catalogue Milani and Knezevic contains much more than 4000 orbits, it is possible to determine more accurate ranges of the proper elements on the basis of the catalogue. The values of the ranges of proper elements, represented by Zappala V. and Cellino A. (1994), were used as reference points to search of asteroid families in coordinates $a' - e'$ and $a' - i'$. Exacter estimations of borders of ranges were established where the concentration of the family asteroid orbits decreases up to concentration of the background asteroid orbits. The estimations of the concentration were carried out by sight.

Table 2 List and rangers of proper elements for asteroid families having more 10 bodies in the IRAS catalogue

| Name | $\Delta a'$ (AU) | $\Delta e'$ | $\Delta i'^o$ | $N_{IRAS}$ |
|---|---|---|---|---|
| 8 Flora | 2.175 - 2.280 | 0.098 - 0.173 | 1.1 - 7.8 | 52 |
| 15 Eunomia | 2.57 - 2.70 | 0.125 - 0.162 | 12.0 - 14.0 | 23 |
| 170 Maria | 2.52 - 2.625 | 0.077 - 0.117 | 14.3 - 15.6 | 19 |
| 145 Adeona | 2.633 - 2.688 | 0.163 - 0.178 | 11.403 - 11.753 | 16 |
| 668 Dora | 2.762 - 2.807 | 0.19 - 0.20 | 7.6 - 8.3 | 13 |
| 158 Koronis | 2.828 - 2.948 | 0.035 - 0.058 | 1.80 - 2.72 | 24 |
| 221 Eos | 2.970 - 3.030 | 0.050 - 0.100 | 9.0 - 11.3 | 100 |
|  | 3.030 - 3.074 | 0.058 - 0.117 | 8.4 - 12.0 | 7 |
| 24 Themis | 3.09 - 3.23 | 0.13 - 0.17 | 0.5 - 2.0 | 94 |

If asteroid orbit falls within the range both in coordinates *a'* - *e'*, and in coordinates *a'* - *i'* of any family, this asteroid belongs to the family. Among asteroids, relating to the families there were selected the bodies included in the IRAS catalogue. The list of the families having more than 10 asteroids included in the IRAS catalogue ($N_{IRAS}$), are represented in Tab. 2. The ranges of the proper elements are shown in the table as well. In this case it is necessary to pay attention to Eos family, which is divided into two sites. The fact is that not for all asteroid families it is possible to determine concrete ranges of proper eccentricities and inclinations. Sometimes these ranges are not constant at all values of the proper semimajor axes of the family. In other words, the space occupied by family in coordinates *a'* - *e'* or/and $a' - i'$ does not form a rectangular. To some families, for example Eos, it is especially visibly. Moreover, this family is evidently divided into two parts along the scale *a'*. It is possible we deal with two different families having close ranges of proper semimajor axes and almost coincident proper inclinations and eccentricities.

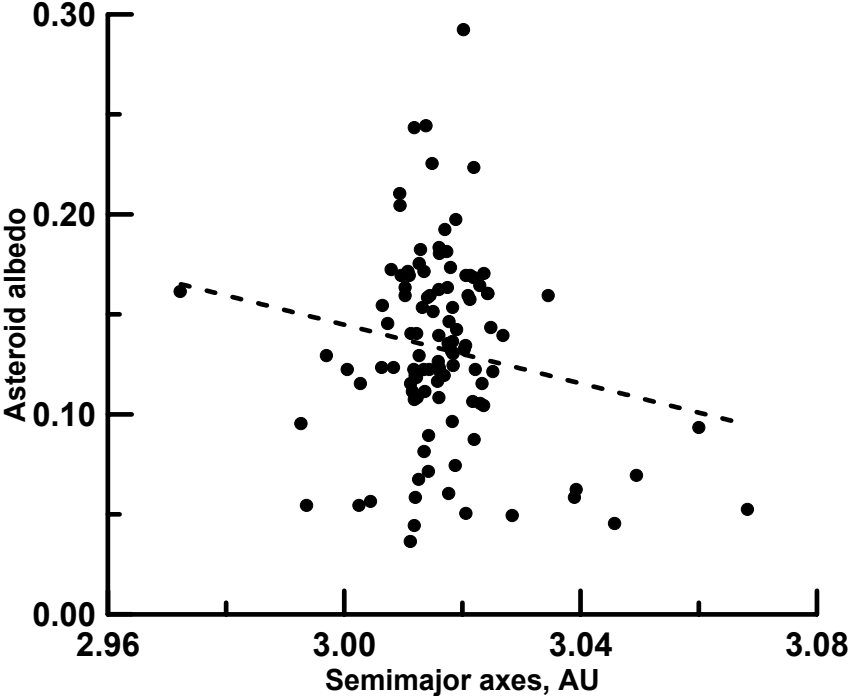

Fig6 Distribution albedo – semimajor axis for Eos family asteroids

To each of these families the dependences of albedo on semimajor axes were constructed and average linear dependences of type (2) were lined. To five families the average dependences have appeared practically parallel to the axis $a'$, i.e. there is no evident connection between of average value of asteroid albedo and semimajor axes. For three families it is revealed a significant decrease of average value of asteroid albedo under semimajor axes increase. For Eos family (Fig. 6) such decrease is significant in according to F test at a level about 0.02. One can see from the figure the average value of asteroid albedo in the right part of the family ($a > 3.03$ AU) is appreciable less than average albedo in the left part. However, such significance can be explained by the fact we deal with two different families. Therefore this dependence we shall not consider as an evident confirmation of action of possible non-gravitational effect.

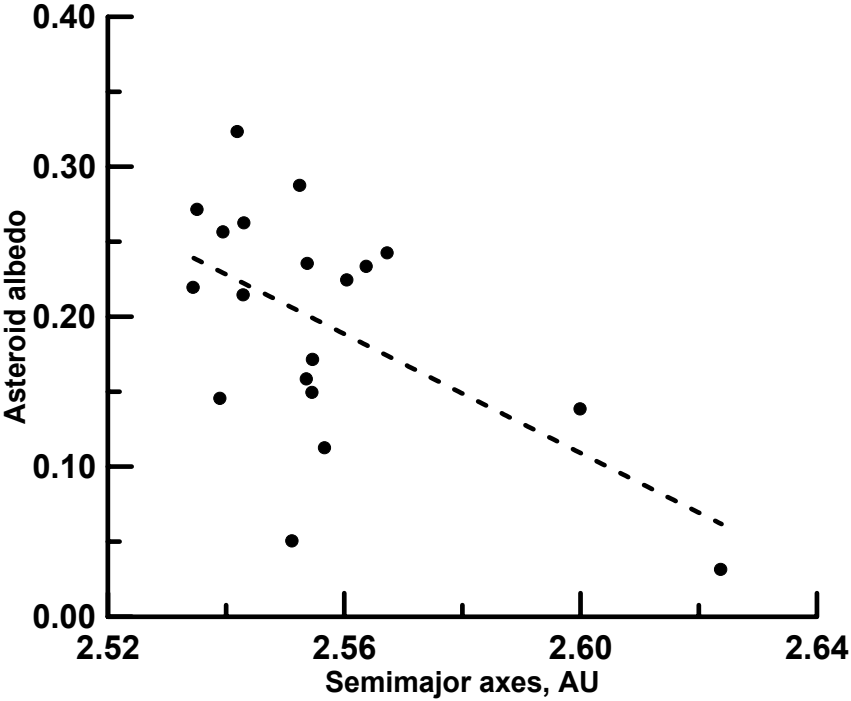

Fig7 Distribution albedo – semimajor axis for Maria family asteroids

The dependence $p(a)$ for Maria family is shown in Fig. 7. The average dependence (dotted line) in according to F test is significant at a level about 0.01. However, one can see, that the principal contribution in the dependence make two orbits with $a' > 2.58$ AU. Therefore the dependence can't be a cogent confirmation of action of a possible non-gravitational effect as well. For more reasoned conclusion touching this dependence it is necessary to fill up the right part of the family with the asteroid albedos.

The dependence $p(a)$ for Flora family is shown in Fig. 8. The average dependence (dotted line) in according to F test is significant at a level about 0.005. The error of factor $b_0$ is 3.6 times less as the factor value. It can seem that three asteroids with albedo greater 0.35, which in Fig. 5 are a little separated from other points, play a vital part in this case. Without the account of these bodies, direction of the regression is unchanged and its significance remains at a level less 0.005.

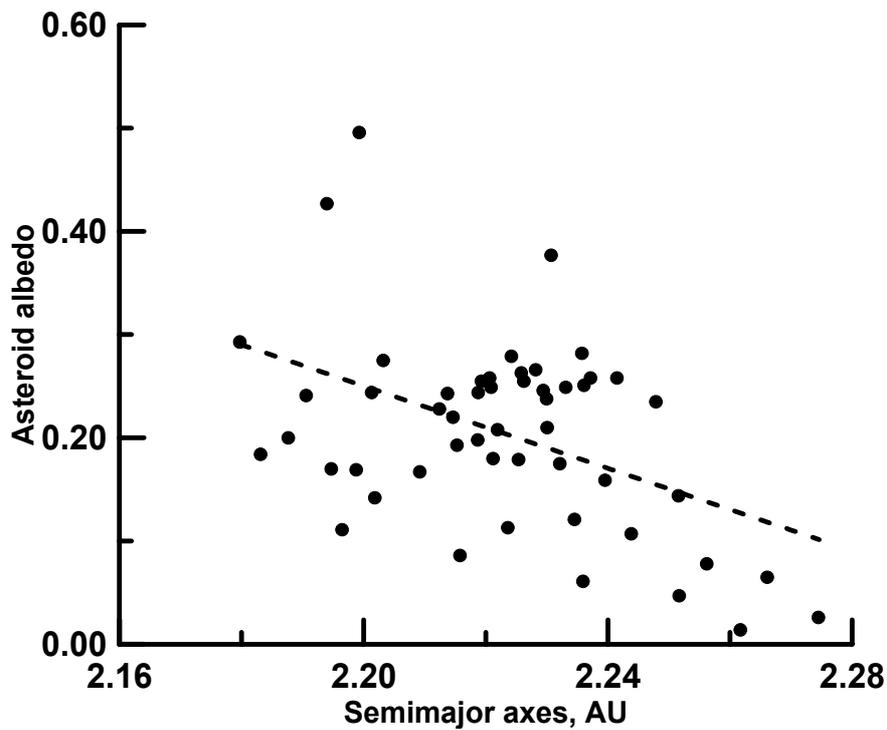

Fig8 Distribution albedo – semimajor axis for Flora family asteroids

In our opinion the distribution *p(a)* for Flora family is a convincing observational confirmation of an existence possibility of non-gravitational effect, which causes spatial separation of asteroids with different albedos. In principle, it is difficult to explain this distribution by other reasons. As semimajor axes aren't subjected by influence of secular perturbations, it is logically to explain the observational dependences by action of some non-gravitational effect. The fact that such confirmation is obtained for Flora asteroid family is quite logical. After all, this family is located most close to the Sun, and action of any non-gravitational force should as much as possible become apparent on the bodies of this family.

It is necessary to say some words about influence of observational selection on distributions *p(a)* for asteroids. At the ordinary observational selection, relative number of observed low-albedo bodies decreases. The observational selection for the IRAS satellite has some specific. The albedos in the IRAS catalogue were determined on the registered own radiation from asteroids on wave-lengths: 12 microns, 25 microns, 60 microns and 100 microns. The own radiation, in one's turn, depends on amount of absorbed solar radiation. Hence, own radiation from low-albedo asteroids $I_1$ will be greater, on the average, than own radiation from high-albedo ones $I_2$. If the ratio $I_1/I_2$ notably increases with semimajor axes increase, the dependences *p(a)*, represented in Figs. 6, 7 and 8, can be to a greater or lesser extent caused by observational selection.

The performed numerical estimations have shown, that the IRAS selection varies the ratio of numbers of bodies with albedo 0.05 and 0.15 no more than by 4% at semimajor axes change from internal edge of the MBA to the external edge (from 2.2 AU up to 3.5 AU). As for separate asteroid families the ranges of semimajor axes are far narrowly, the maximal influence of such selection makes 0.3%. Therefore could say, that the observational IRAS selection by no means influence to the dependences *p(a)*, represented in Figs. 6, 7 and 8.

We understand that the dependences *p(a)* represented in Figs. 6 and 7 aren't as evident confirmations of action of possible non-gravitational effect. They are presented as accessory arguments. There are only three asteroid family for which dependence *da(p)* is statistically significant, and all three dependences are qualitatively similar. For more well-founded conclusion on a reality of similar non-gravitational effect it is necessary to have more complete asteroid samples of greater number of families.

There is a rather simple way to confirm or reject the conclusion of existence of the non-gravitational effect in near future. If, at replenishment of asteroid albeods database, distributions *p(a)* for the majority of the asteroid families will be obtained like for Flora family, it will mean that the non-gravitational effect is real.

## 6. A possible physical mechanism causing such non-gravitational effect

It is clear, reality of such non-gravitational effect requires the physical description of its mechanism as well. A detailed description of such mechanism needs a separate research that is planned further. In it we can only point to a fundamental possibility. Very likely the Yarkovsky effect can't serve an explanation of such orbit changes for so large bodies. In the majority of papers it is considered, that the Yarkosvky effect have action only upon sub-kilometer bodies and less. Though in some papers (Bottke William F. and Vokrouhlický David 2001), to explain semimajor axes drift of asteroid families, the influence of this effect on bodies up to 20km is supposed.

In our paper, semimajor axes drift of larger bodies is being discussed. Therefore it is possible to assume action of non-gravitational force like for comet. Probably, there is some quantity of volatile substances inside large asteroids. After the asteroid disruption these substances appear on the surfaces of splinters. At action of solar radiation, the volatile substances will come off the splinters. The additional impulse, obtained by a splinter, will be a few orders greater than at over-radiating of the absorbed radiation.

The bodies with appreciably different albedos have different physical properties of surfaces as well. More dim surfaces will warm up faster to temperature, sufficient for evaporation of volatile substances. At the prograde rotation and with certain periods, the evaporation from more dim asteroids can, basically, occur in the opposite direction to orbital motion. And for bright bodies – along the lines of the motion. Hence, semimajor axes of orbits of more dim bodies will on average increase, and of more bright – will decrease.

It is clear that evaporation of volatile substances from an asteroid surface occurs not just along the lines of the orbital motion (or in the opposite direction). The evaporation should occur in a wide range of directions. Therefore, the above opinion, that chaotic changes of asteroid space position, caused by non-gravitational effect, may be grater than systematic ones and may affect residuals, is quite logical.

To spatial separation of bodies with different albedos, it is necessary non-isotropic distribution of the rotation periods and spin obliquities of asteroids. These conditions are not very unreal. So, according to (Donnison J.R. 2003), asteroids with diameters $D > 33$ kms have a single Maxwellian distribution on rotation periods with the evident maximum with the average period 11.4 hours. Harris, Alan W. and Pravec, Petr (2006) point that non-random spin axes orientation of asteroids with diameters to a few tens of km can be caused by solar radiation pressure. It is especially interesting, that non-random axes orientation is inherent in separate asteroid families. Such feature of family asteroids can be explained by Yarkovsky effect (Nesvorný David and Bottke William F. 2004).

It is clear, that those arguments do not explain the physical mechanism of the possible non-gravitational effect, and only outline probable ways for its explanation.

### 7. Influence of possible non-gravitational effect on NEAs population

Supposition of existence possibility of non-gravitational effect may cause a natural question of influence such effect on NEAs numbering. It is clear the effect must promote increase in near-Earth asteroids. Mechanism of the increase consists in transfer of asteroid orbits to resonance zones with Jupiter and Saturn under the influence of the effect. It is recognized, the NEAs population is replenished just from the resonances in MBA. Therefore, it is necessary to clear up if contradicts this supposition with observed numbering of near-Earth asteroids.

The orbital resonances with Jupiter 3 : 1 and 5 : 2, and secular resonance with Saturn $v_6$ as well are consider as the main sources of NEAs. As the assumed non-gravitational effect should change semimajor axes of asteroid orbits, therefore the orbital resonances should be replenished most of all. Speed of such replenishment can be approximately estimated with the help of aforecited dependences $da(p)$. It was obtained that $da = 0$ at $p \approx 0.10$ practically for all dependences. For asteroids with less albedos the semimajor axes should increase, and at larger albedos – should decrease.

As stated above, separation speed for the most dim and the most bright asteroids with the sizes 10 - 50km is about $1\times10^{-8}$ AU per year. However, there are very few asteroids with such extreme albedos. Distribution asteroid numbering $N$ on albedo is presented in Fig. 9. One can see a clear maxima at $p = 0.05$. The average values $da$ at that albedo is approximately equal $+1\times10^{-8}$ AU. For asteroids with semimajor axes decrease ($p > 0.1$) a local maxima in the distribution $N(p)$ corresponds to $p \approx 0.15$. At that albedo the average values $da$ is approximately equal $-1\times10^{-8}$ AU. Thus, both the average velocity of semimajor axes increase (for asteroid with low albedos), and the average velocity of semimajor axes decrease (for asteroid with high albedos) are equal about $1\times10^{-9}$ AU per year.

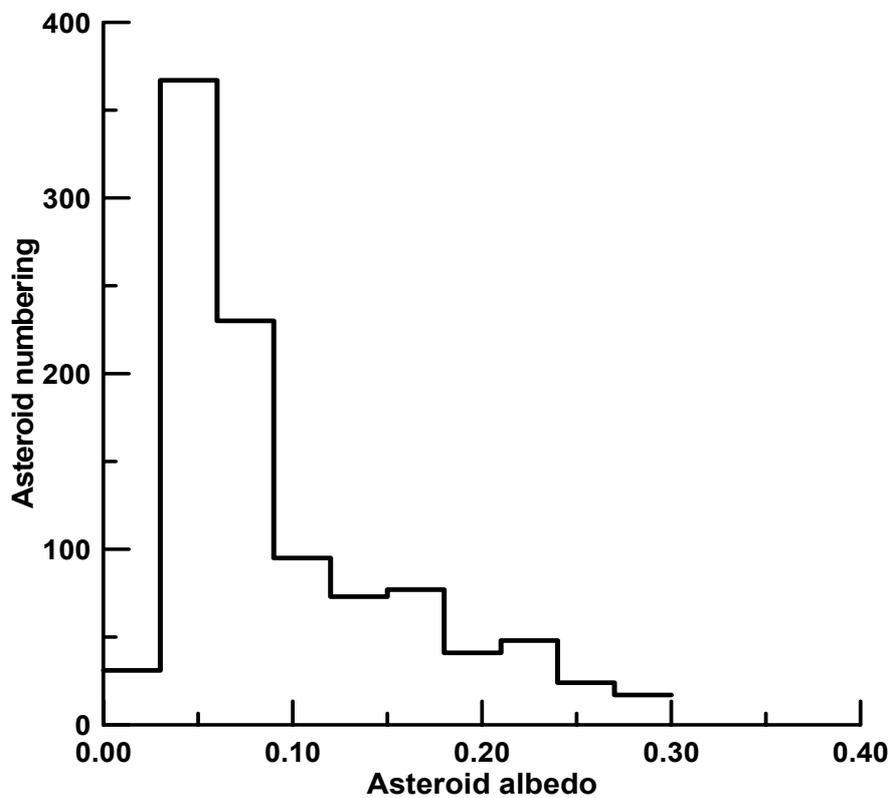

Fig9 Distribution of MBAs numbering on albedo

Besides, it is necessary to estimate influence of the non-gravitational effect to replenishment of smaller near-Earth asteroids. For this purpose the asteroid sizes from 1 to 5км were chosen. It was accepted at the same time, that change speeds of semimajor axes for bodies of from 1 to 5km ten times exceed the correspondent values for bodies of 10–50км. The impulse of ejecting matter is proportional to surface area of the asteroid ($D^2$), and the asteroid mass is proportional to $D^3$. Consequently, total additional velocity is proportional to $D^{-1}$.

To estimate of replenishment rate for NEAs it is important to know the typical lifetime for such asteroids. According to Foschini, L. et. al. (2000) lifetime for NEAs is about $10^7$ y. Morbidelli, A. (2001) asserts that the dynamical lifetime of the NEAs one order of magnitude shorter than previously believed (i.e. $10^6$ y.). For NEAs with aphelion distances Q > 3.5 AU the lifetime should be even shorter – $10^5$ years (Fernández, Julio A. et. al. 2002). For our task we may used the average lifetime – $10^6$ y.

To a more clear answer our problem, evolution of some hundreds model orbits located near the commensurabilities 3:1 and 5:2, were calculated. The integration intervals amount to a few million years. Perturbations of eight planets (Mercury – Neptune) were taken into account. Besides, semimajor axes of orbits artificially changed in every step by a value corresponds to rate of change of *a* caused by non-gravitational effect.

The calculation results have shown that about 80% of bodies near resonance 3:1 transfer to orbits with perihelion distances *q* < 1.3 AU. The average duration of stay in such orbits is about 500 thousands years. There are about 50% such bodies near resonance 5:2, and the average duration of stay – approximately 150 thousands years. The obtained estimations of duration are very close to ones presented in (Morbidelli, A. 2001, Fernández, Julio A. et. al. 2002).

It is obviously, the NEAs numbering should be about the asteroid numbering, located in zones of semimajor axes, near by the resonances. The dimensions of the zones should be equal values of asteroid drifting during the NEAs lifetime.

In contrast to the orbital resonances, in order to come into secular resonance $\nu_6$ asteroid orbits must have not only the certain values of semimajor axes, but the certain values of inclination as well. The zone of this resonance in coordinates *a – i* is so complicate. There are very few asteroid orbits in MPC catalogue, which can come into this resonance at semimajor axes decrease. According to our estimations, the total number of bodies with $H < 18^m$, which can transfer to NEAs through resonance $\nu_6$ due to non-gravitational effect amounts to 50.

Taking into account the numbering of discovered asteroids near the resonances and observational selection effect, the following result were obtained.

The NEAs numbering of from 10 to 50km, replenished by the effect should amount 3-4 asteroids. There are 5 bodies of from 10 to 50km ($H < 13^m$) among the NEAs for today.

The NEAs number of from 1 to 5km replenished by the effect, should amount about 600. There are about 900 NEAs larger than 1km ($H < 18^m$) in the MPC catalogue to date. The total NEAs numbering should be still more.

Consequently, it is possible to draw a conclusion that supposition on existence possibility of non-gravitational effect doesn't contradict with observed number of near-Earth asteroids.

**Conclusions**

The carried out numerical calculations of asteroid orbits included in the IRAS catalogue and distributions "albedo – semimajor axes" for asteroid families indicate to an existence possibility of non-gravitational effect, causing spatial separation of asteroid with different albedos. Semimajor axes of low-albodo asteroids are, on average, increasing as compared with high-albedo ones.

An analysis of errors and residuals in the asteroid catalogues is evidence of such supposition.

For the definitive conclusion on a reality of such non-gravitational effect and for more exact numerical estimations it is necessary additional observed data and theoretic researches.

**Acknowledgments.** The author is grateful to prof. Lupishko D.F. (Scientific Research Institute of Astronomy of V.N. Karazin Kharkov National University) for useful advises and for given references, and to prof. Shor V.A. (Institute of Applied Astronomy, St. Petersburg) for rendering necessary data and explanations.